\documentclass[pra,twocolumn,showpacs,amsmath,amssymb]{revtex4}
\usepackage{graphicx}% Include figure files
\usepackage{dcolumn}% Align table columns on decimal point
\usepackage{bm}% bold math
\usepackage{subfigure}
\usepackage{amsmath}
\usepackage{amssymb}
\usepackage{color}
\usepackage{siunitx}  %for tables with decimal points very usefull for clocks ...
\def\prt{/\sqrt{\tau}}

\makeatother

\begin{document}
\title{A transportable optical lattice clock with $7\times10^{-17}$ uncertainty}

\author{S. B. Koller, J. Grotti, St. Vogt, A. Al-Masoudi, S. D\"orscher, S. H\"afner, U. Sterr, Ch. Lisdat}
\affiliation{Physikalisch-Technische Bundesanstalt, Bundesallee 100, 38116 Braunschweig, Germany}

\date{\today}

\begin{abstract}
We present a transportable optical clock (TOC) with $^{87}$Sr. Its complete characterization against a stationary lattice clock resulted in a systematic uncertainty of ${7.4 \times 10^{-17}}$ which is currently limited by the statistics of the determination of the residual lattice light shift. 
The measurements confirm that the systematic uncertainty is reduceable to below the design goal of $1 \times 10^{-17}$. 
The instability of our TOC is $1.3 \times 10^{-15} \prt$.
Both, the systematic uncertainty and the instability are to our best knowledge currently the best achieved with any type of transportable clock.
For autonomous operation the TOC is installed in an air-conditioned car-trailer. 
It is suitable for chronometric leveling with sub-meter resolution as well as 
intercontinental cross-linking of optical clocks, which is essential for a redefiniton of the SI second.
%for intercontinental frequency comparisons at low uncertainty and instability, which are essential to prepare for a redefinition of the SI second.
In addition, the TOC will be used for high precision experiments for fundamental science that are commonly tied to precise frequency measurements and it is a first step to space borne optical clocks.
%To these ends, the clock is installed in an air-conditioned trailer in which it can be operated. 
\end{abstract}

\maketitle

The best clocks in the world reach a fractional systematic uncertainty at the low $10^{-18}$ level \cite{nic15,ush15,hun16,sch16} and instabilities near or even below $10^{-16}\prt$ \cite{sch16, ush15, alm15, nic15, hin13}, surpassing the clocks realizing the SI second in both by two orders of magnitude. 
This has triggered a discussion about a redefinition of the SI second \cite{rie15, mar13b}, pushes the frontiers of precision spectroscopy and tests fundamental physics \cite{nem16,gre16,let13,hun14,god14,for07}, and enables chronometric leveling \cite{ver83,bje85,bje86,lis16,tak16}, where gravitational redshifts are exploited to measure height differences.

So far, the operation of optical clocks has been constrained to laboratories. 
However, transportable clocks are required for the necessary flexibility in the choice of measurement sites for applications like chronometric leveling.  
Also, they are highly interesting for frequency metrology and time keeping in creating a consistent worldwide network of the next-generation ultraprecise clocks.
Although comparisons at the full performance level of state-of-the-art optical clocks are possible on a continental scale \cite{lis16, tak16} through a few specialized optical fiber links \cite{pre12, rau15, chi15}, intercontinental links are so far restricted to satellite-based methods that cannot fully exploit the clock performance \cite{hac14}. 
A transfer standard enables world-wide interconnections between optical clocks and will thus benefit the efforts towards a redefinition of the SI second.

Making laboratory setups compact and robust for transport is also the first step towards granting a wide community of users access to these devices \cite{nie00, fis04, par11}. 
Furthermore, transportability is a first step towards applications of optical clocks in space. 
Developments in these directions are ongoing for optical lattice clocks (OLCs) with strontium \cite{bon15,pol14}; however, to our knowledge the single-ion clock reported recently \cite{cao16}  is the only other transportable clock with uncertainty below $10^{-16}$.

The requirements on such a TOC are challenging indeed:
To enable comparisons of other optical clocks it has to achieve uncertainties similar to those of the clocks to be tested or at least considerably better than what can be reached by comparing to primary cesium clocks \cite{gre16,lod16,fal14,let13}. 
Further, for geodetic applications, i.e., chronometric leveling, a resolution of below ten centimeters is required to compete with established methods that connect sites separated by a few hundreds of kilometers.
%Precision clocks measure this by locking at the fractional difference that is equal the to the gravity potential difference divided by the square of the speed of light.
This means that fractional gravitational red shifts of $10^{-17}$ and below must be resolved by the TOC.
Preferably, the frequency instability should be in the range of typical optical clocks to enable reasonably short measurement times.
%Since the fractional shift $\Delta \nu / \nu$ of the clock frequency $\nu$ related to the gravity potential difference $\Delta U$ between the two clocks is \cite{Str04}
%
%\begin{equation}
%  \frac{\Delta\nu}{\nu}=\frac{\Delta U}{c^2}
%	\label{eq:red_shift}
%\end{equation}
%
%with $c$ the speed of light, fractional shifts of $10^{-17}$ and below must be resolved by the transportable clock.
 
\begin{figure}[ht]
	\includegraphics[width=0.45\textwidth]{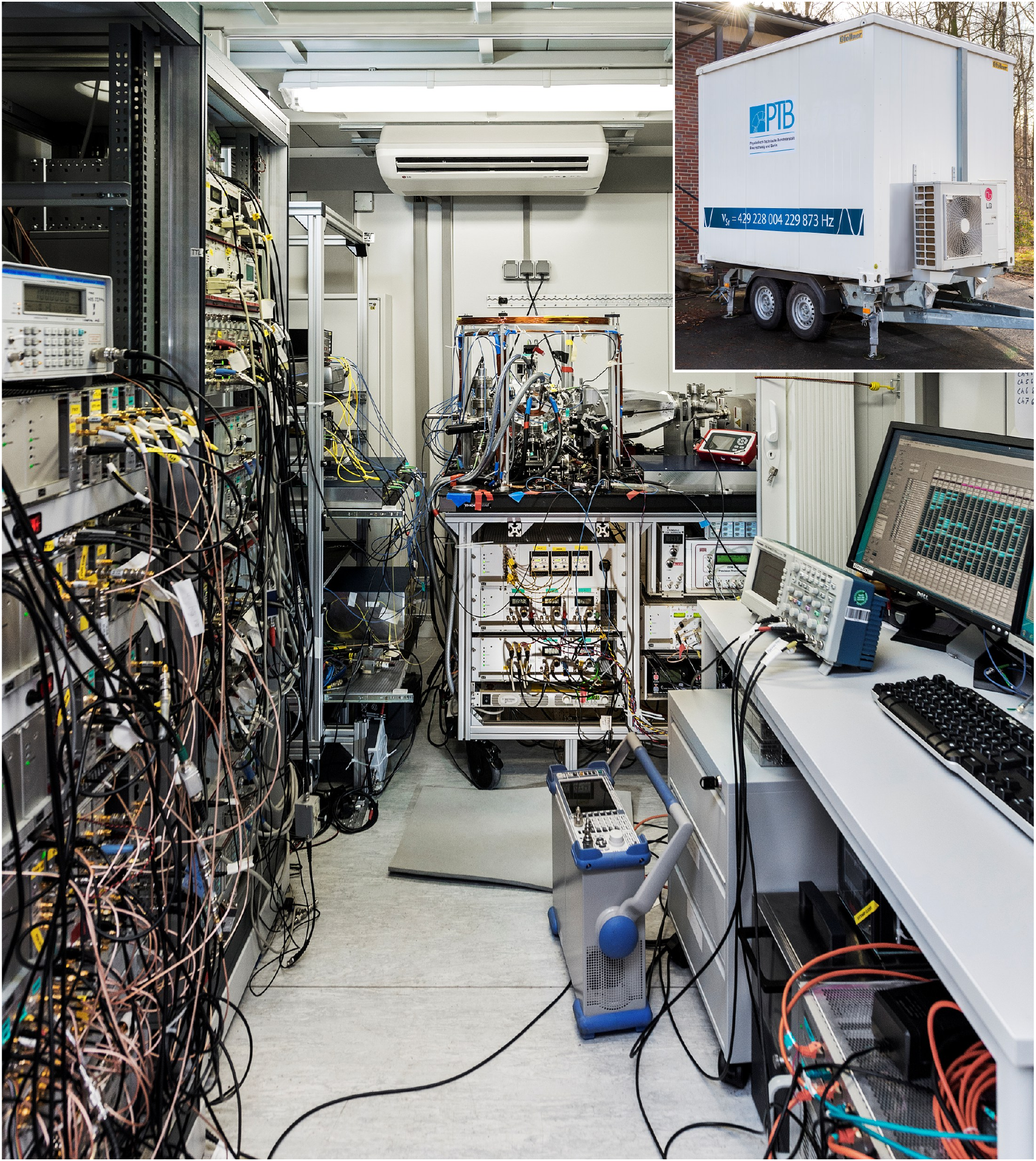}
	\caption{\label{fig:container}(color online) View into the car-trailer for transport and operation. 
	Front, left: electronics for the laser systems. 
	Back, left: laser systems for cooling and trapping, and the ULE cavities to lock the lasers.
  Back center: physics package.
  Front right: computer control.
	Not shown are the interrogation and lattice laser setups. %(see text for further information)
The interior dimensions of the container are 2.2~m $ \times$ 3~m $\times $ 2.2~m. The  mass of the depicted experimental setup is approximately 800~kg.
        Inset: the car-trailer from outside.}  %The container with cart less than 3000~kg, where the experimental setup itself has a mass of approximately 500~kg.}
\end{figure}

These requirements are considerably beyond the properties achieved with the best transportable microwave clock \cite{biz04}.
For this atomic cesium fountain clock an uncertainty of $5.9\times10^{-16}$ \cite{abg12} and an instability of $1.8 \times 10^{-13}\prt$ \cite{gue12a} have been reported and exploited in several campaigns \cite{nie00, fis04, par11}.
%{\bf I haven't found the original numbers in a paper for the stability}.
The $^{40}\mathrm{Ca}^+$-ion clock reported in \cite{cao16} reaches an uncertainty of $7.7 \times 10^{-17}$ and an instability of $2.3 \times 10^{-14}\prt$, but has not left the laboratory yet.
Here we present a transportable OLC that is characterized and compared to an established, stationary optical frequency standard \cite{fal14, alm15, gre16, lis16} and tested outside the laboratory in a transportable car-trailer (Fig.~\ref{fig:container}). 

This clock uses the $ (5s^2) {^1S_0}-(5s5p) {^3P_0}$ transition in $^{87}\mathrm{Sr}$ at 698 nm as reference transition, which is interrogated in atoms laser-cooled into and confined in a one-dimensional optical lattice. 
The OLC comprises four main parts: The physics package, where the atoms are prepared and the reference transition is interrogated; the laser systems for laser cooling, state preparation, and trapping; a highly frequency-stable interrogation laser system; and the computer control to generate the experimental sequence and feedback to the interrogation laser frequency.
For a TOC it is essential that these parts are compact in size and robust in construction to provide fast and reliable measurements at different locations. 

Our cooling and preparation laser systems use commercial diode lasers integrated into five compact breadboards with half-inch optics \cite{vog15,vog16}.
These breadboards contain acousto-optical modulators (AOMs) and optical shutters for light switching and frequency modulation.
They are connected to the physics package with polarization maintaining fibers.
The boards have a size of 30~cm $\times$ 45~cm $\times$ 6~cm and a mass of less than 8~kg each. 
Since we do not aim for extreme compactness, we still use standard control electronics.

Similarly, the compact physics package is mounted on a 120~cm $\times$ 90~cm optical breadboard, which has not been miniaturized to avoid trade-offs with the clock performance.
The Zeeman slower for efficiently loading atoms into a magneto-optical trap (MOT) is based on permanent magnets to avoid heat load from a solenoid near the interrogation region of the atoms \cite{vog16}.
Given the small volume available for a transportable setup, this helps preventing thermal inhomogeneity of the environment of the atoms as the blackbody radiation (BBR) shift is typically the largest source of uncertainty in Sr OLCs that are operated in a room-temperature environment \cite{nic15,blo14}.
The physics package and laser-systems have been tested successfully in the car-trailer, achieving similar atom numbers and temperatures as in the laboratory.

The reference cavity to pre-stabilize the frequency of the interrogation laser is a highly critical part of the TOC. 
Rigid mounting \cite{hae15b,ste11b} is employed in our reference cavity, assembly as standard soft supports (e.g. \cite{hae15a}) would not withstand transport.
This laser system is a further development of an earlier cavity-stabilized laser system \cite{vog11} and reaches a frequency instability flicker floor of about $4 \times 10^{-16}$ after 10~s.
For best performance in a measurement the 12~cm long reference cavity and the interrogation laser are placed outside the transport container in a seismically quieter environment.
%
%\begin{figure}
%	\def\svgwidth{0.45\textwidth}
%	\input{clockprot.pdf_tex}
%	\caption{\label{fig:clockprot}(color online) Typical measurement protocol of the TOC.
%	The relative width of the blocks are proportional to the total time. Blue: first stage cooling, clean-up, detection (461~nm, ${^1S_0}-{^1P_1}$ transition); red: second stage cooling, spin polarization (689~nm, ${^1S_0}-{^3P_1}$ transition); dark red: 698~nm reference transition interrogation; orange: repumping (679~nm and 707~nm, ${^3P} - {^3S}$ transitions). Not described blanks show delays that serve to close shutters or settle magnetic fields.}
%\end{figure}

The measurement sequence is comparable to other $^{87}\mathrm{Sr}$ OLCs.
Typically, the cycle time is 900~ms with a duty factor of 0.16:
A rotary atom shutter is used to shield the atoms during the interrogation from the BBR of the oven   at 500~$^\circ$C loading the MOT. 
Opening the shutter takes 10~ms. 
However, we wait for additional 90~ms which increases the stability of the atom number in the lattice. 
%the mechanical damping of the opening process seems to need more time than the preparation of the first-stage magneto-optical trapping (MOT) (see below) and therefore 100~ms are reserved for this operation. 
The first MOT loading stage (300~ms) on the 461~nm ${^1S_0}-{^1P_1}$ transition is followed by a second cooling stage using the 689~nm ${^1S_0}-{^3P_1}$ transition, which is split into a  broadband red MOT (80~ms) and a single-frequency MOT (80~ms), during which the optical lattice is loaded.
The lattice is tilted by approximately 50$^\circ$ against gravity and operated at the Stark shift cancellation wavelength near 813~nm \cite{kat02}.
The full trap depth is typically about 80~$E_{\rm r}$, where $E_{\rm r}$ is the lattice photon recoil energy.
The atoms trapped in the lattice are spin-polarized alternatingly to one of the stretched state ($|m_F|=9/2$) of the ground state manifold (30~ms).
To remove the atoms in other Zeeman levels (clean-up), a $\pi$-pulse is driving the atoms in the chosen Zeeman state on a resonant $\pi$-transition in a magnetic field of about 1.9~mT to the $^3P_0$ state (30~ms).
Atoms remaining in the $^1S_0$ manifold are expelled from the lattice by a pulse of 461~nm light (10~ms).
The actual interrogation is performed in a magnetic field of about 45~$\mu$T parallel to the linear polarization of the lattice with a Rabi $\pi$-pulse duration in the range of 100~ms to 150~ms.
We use a normalized electron-shelving detection technique by applying a combination of the 461~nm fluorescence detection with 707~nm and 679~nm light to repump the atoms to the ground state, which determines the excitation probability at the frequency of the interrogation laser \cite{fal11}. 

\begin{figure}
    \centering
	\includegraphics[width=0.4\textwidth]{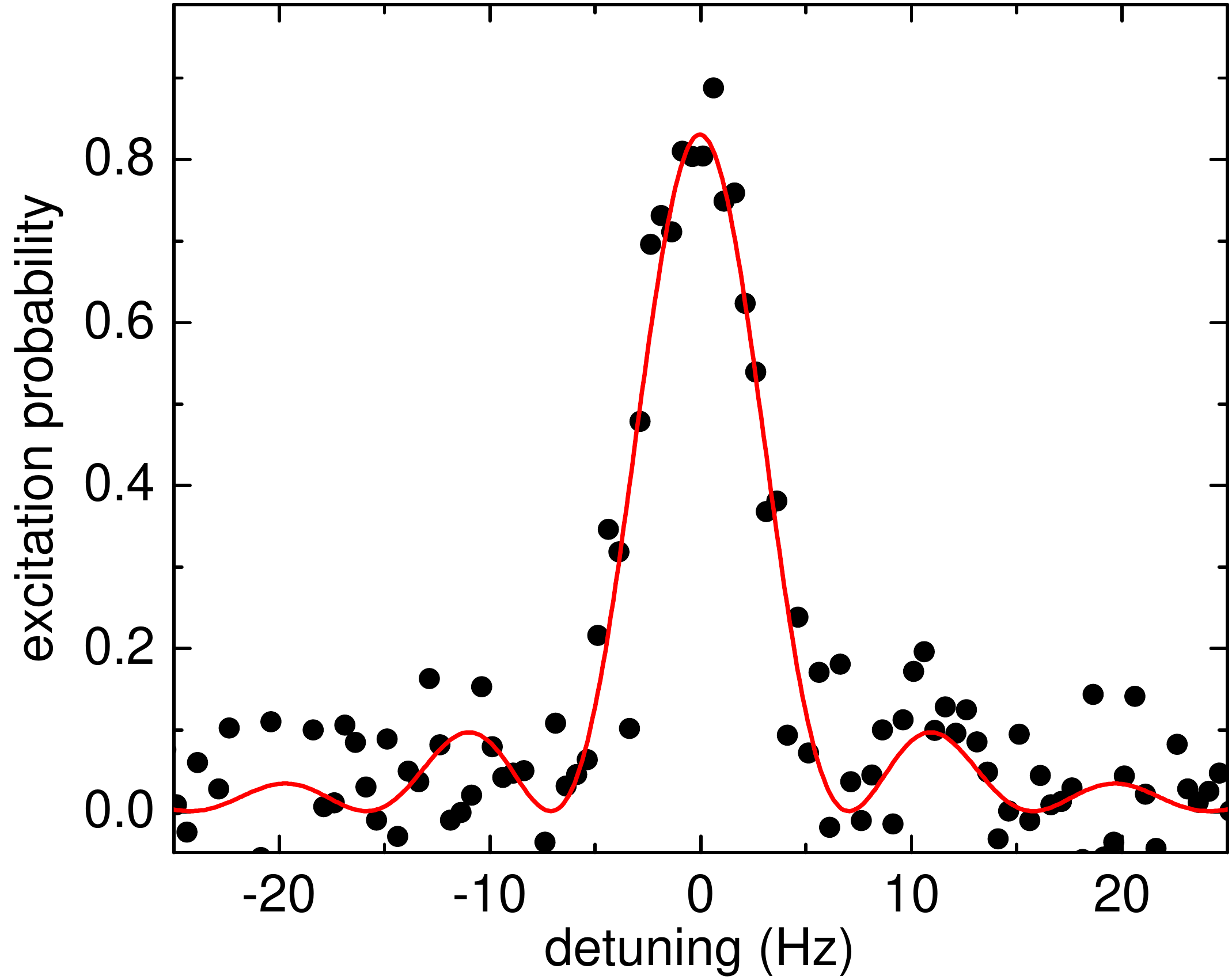} %O:\4-3\4-3-Alle\Papers\In_Arbeit\Transclock_PRL2016\Figs
	\caption{\label{fig:trans_line}(color online) 
	%The solid line indicates an instability of $1.3 \times 10^{-15}\prt$.
	Scan over the 698~nm reference transition in $^{87}$Sr (dots). 
    The resolved linewidth is 7~Hz, determined by the excitation pulse length of 112~ms. 
    The solid line shows a corresponding Rabi excitation line profile.}
\end{figure}

We apply this technique on both half-width points of each $\pm 9/2$ transition, addressing them and pulsing the light by using an AOM.
%After four cycles, the computer program combines the resulting excitation probabilities at the different frequencies to determine the frequencies for the next four cycles including the linear Zeeman splitting and the drift of the ULE cavity.  
A computer program evaluates these four interrogations and tracks the center frequency, the linear Zeeman splitting, and drift of the reference cavity.
In this way it stabilizes the interrogation laser to the linear-Zeeman-shift-free transition frequency.
Tuning the interrogation laser frequency across a single Zeeman transition, we observe a Fourier-limited linewidth of 7~Hz as shown in Fig.~\ref{fig:trans_line}.

\begin{figure}
    \centering
	\includegraphics[width=0.5\textwidth]{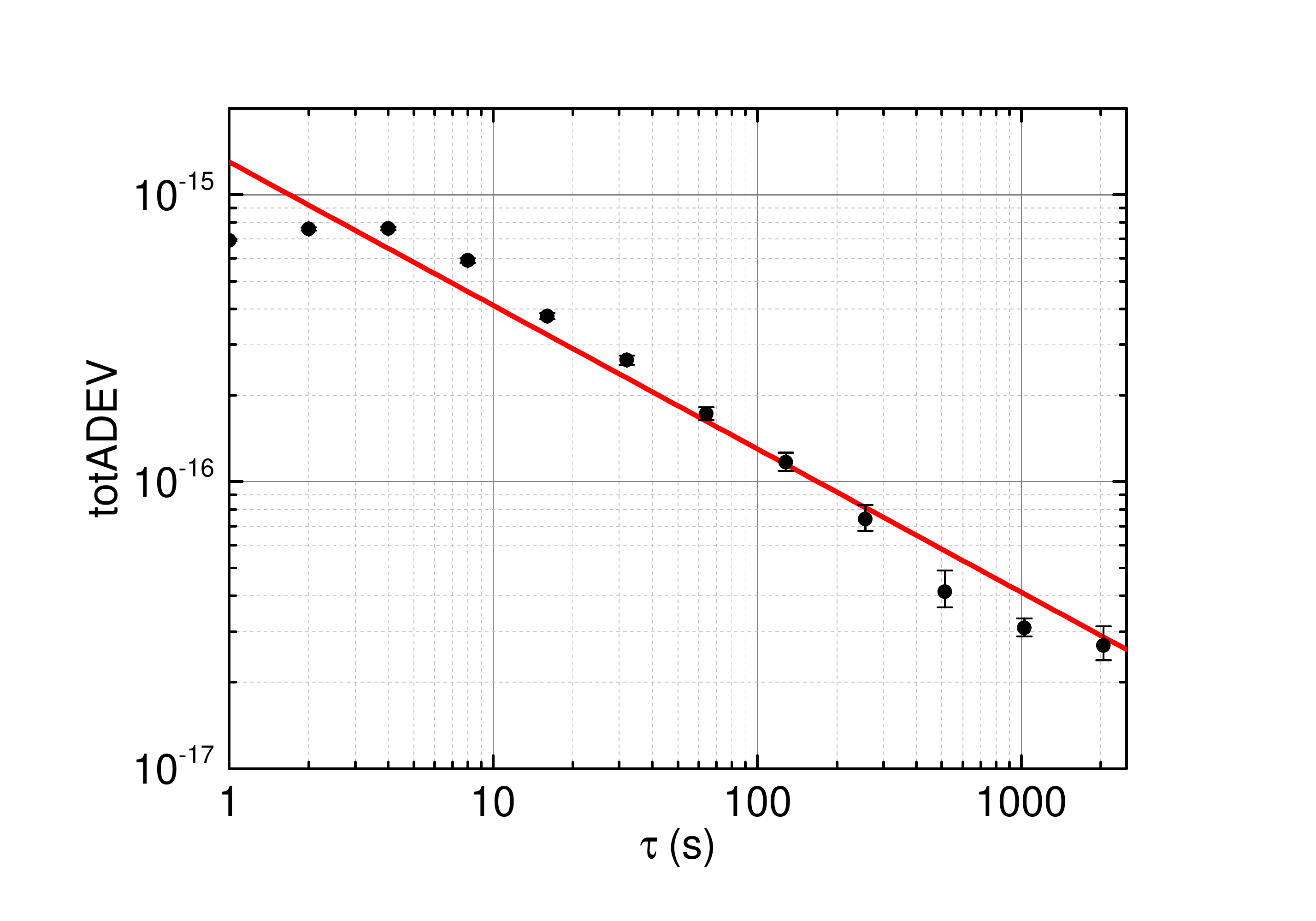} %O:\4-3\4-3-Alle\Papers\In_Arbeit\Transclock_PRL2016\Figs
	\caption{\label{fig:lw_ro_Allan}(color online) Total Allan deviation (totADEV) of the frequency ratio $\nu_{\rm trans}/\nu_{\rm stat}$ (dots).
	The solid line indicates an instability of $1.3 \times 10^{-15}\prt$.}
	%Inset: Scan over the 698~nm reference transition in $^{87}$Sr (dots). 
    %The resolved linewidth is 7~Hz, determined by the excitation pulse length of 112~ms. 
    %The solid line shows a corresponding Rabi excitation line profile.}
\end{figure}

To evaluate the TOC, we perform a direct comparison with our stationary $^{87}$Sr OLC at PTB \cite{fal14,alm15,gre16}. 
The beat frequency between the two reference lasers is recorded with a dead-time-free counter.
No elements of the two clocks are shared such that the two clocks are fully independent.
The instability of their frequency ratio, $\nu_{\rm trans}/\nu_{\rm stat}$, is expressed by the total Allan deviation (Fig.~\ref{fig:lw_ro_Allan}).
An instability of about $1.3 \times 10^{-15}\prt$ is observed, which is governed by the transportable clock, since the stationary system exhibits an instability in the low $10^{-16}\prt$ range \cite{alm15}.
The instability of the TOC is dominated by the performance of the interrogation laser \cite{hae15b} through the Dick effect  \cite{dic87}.
This instability is lower than achieved in typical single ion clocks \cite{cho10,god14,cao16,hun16}, and comparable to many high perfromance lattice clocks \cite{let13,lod16,hac15,tak11}.

%This instability is less than on order of magnitude worse than the most stable single optical lattice clocks \cite{let12, nic15, alm15}, better than typical single ion clocks \cite{ros08,cho10,god14,fal14,cao16}, and comparable to high performance lattice clocks \cite{let13,lod16,hac15,cao16}.

Beyond the instability of the transportable clock, the ratio measurements also test the agreement of both Sr OLCs within their uncertainties.
The uncertainty of the stationary clock has been evaluated repeatedly and compared  with primary Cs clocks \cite{gre16,fal14} and another Sr OLC down to a fractional uncertainty of $5 \times 10^{-17}$ \cite{lis16}.
%Therefore, the measurements discussed here are a stringent test of the TOC.

\begin{figure}[t]
	\includegraphics[width=0.5\textwidth]{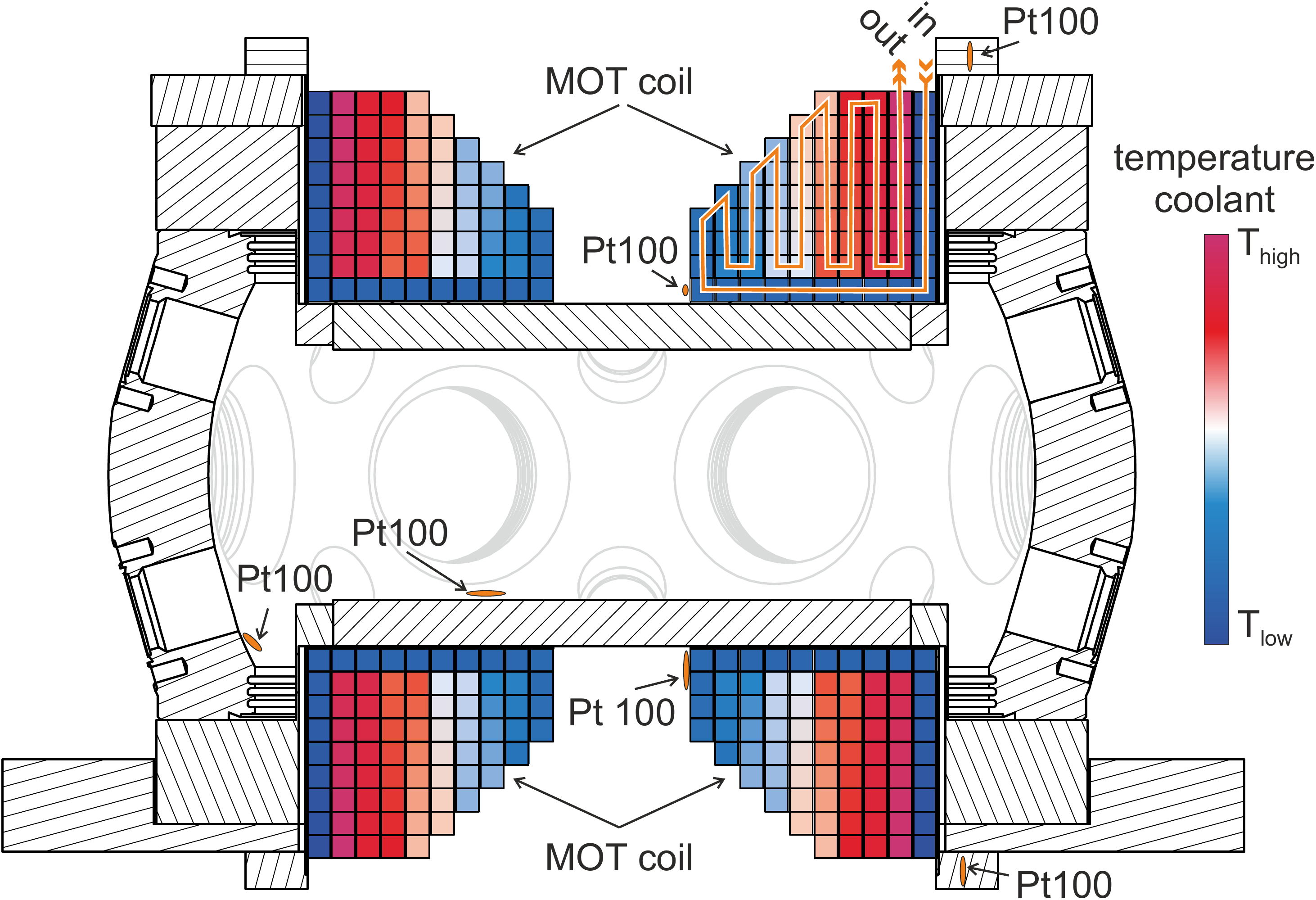} 
	\caption{\label{fig:coils}(color online) A section through the main vacuum chamber. The MOT coils consisting of hollow, square wires with the coolant in the bore are located in re-entrant flanges. The outline is indicated by the coolant flow and the color-coded coolant temperature. The windings exposed to the chamber are cooled first and are thus temperature controlled best. The locations of some Pt100 temperature sensors are also indicated.}
\end{figure}

The uncertainty of the TOC was evaluated along the same lines as discussed for our stationary system \cite{fal12,gre16}:
The BBR shift causes the largest fractional correction of about $5 \times 10^{-15}$.
To ensure a precise control of this shift even in a less temperature-stable environment than a laboratory, constructional details of the physics package are important.
We avoid excessive power dissipation in the physics package by using the Zeeman slower mentioned above.
For the coils generating the MOT and bias magnetic fields, we opted for compact coils with efficient water cooling (Fig.~\ref{fig:coils}).
The cooling water temperature is stabilized to better than 100~mK by a thermostat, which not only removes the energy dissipated in the coils, but effectively stabilizes the temperature of a large part of the environment of the atoms.
Further, by adjusting the temperature of the coolant we minimize the temperature difference of the warmest ($T_{\rm max}$) and coolest ($T_{\rm min}$) spots of the vacuum chamber.
The temperature is measured at eight locations out- and inside the vacuum chamber by platinum resistance sensors with an uncertainty of 40~mK, where we have taken care to cover the coolest and warmest spots.
According to \cite{gum95} we use $\overline{T} = (T_{\rm max} + T_{\rm min})/2$ as representative temperature to calculate the BBR shift.
 We assign an uncertainty of $(T_{\rm max} - T_{\rm min})/\sqrt{12}$ as we assume the {\lq true\rq}  temperature to lie with constant probability in the interval $[T_{\rm max},T_{\rm min}]$.
Typically, we observe $\Delta T \approx 0.4$~K resulting in an uncertainty contribution of about $9 \times 10^{-18}$ \cite{nic15,mid12}.

Scalar and tensor lattice light shifts \cite{wes11} are determined by monitoring the difference in frequency offsets between the TOC and the stationary clock when operating the former with thermally averaged lattice depths of 50.8~$E_{\rm r}$ and 88.8~$E_{\rm r}$.
The atomic temperature in the lattice is derived from sideband spectra \cite{bla09a} and typically about 3.5~$\mu$K.
Higher-order light shifts are calculated using the coefficients given in \cite{wes11,let13}.
Further details on the lattice light shifts are given below.

The cold-collision shift is determined similarly by varying the atom number in the lattice.
Background gas collision shifts are calculated based on the theory in \cite{gib13}, a lattice lifetime of 3~s and coefficients from \cite{mit10a}.
While the linear Zeeman shift is directly removed by the stabilization protocol, the quadratic Zeeman shift has to be corrected independently.
This is straightforward as the splitting of the Zeeman transitions provides a direct measure of the magnetic field sampled by the atoms.
Tunneling between lattice sites is strongly suppressed due to the tilted and deep lattice \cite{lem05}.
The inner surfaces of the re-entrant flanges, which are the surfaces closest to the atoms (Fig.~\ref{fig:coils}), are coated with a conducting material (Indium Tin Oxide) on top of an anti-reflective coating.
With a separation of the windows of 54~mm, possible patch potentials of up to 100~mV, and the coefficient from \cite{mid12}, we estimate a maximum DC Stark shift far below $1 \times 10^{-18}$.
Other known uncertainties from servo errors, optical path length variations \cite{fal12}, and line pulling are below $10^{-18}$ and will not be discussed here.

\begin{table}[t!] %the table is the second table of the excel file SR1_SR2_comparison_with_lattice@35MHz.xlsx
	\begin{tabular}{lS[table-format=3.2]S[table-format=1.2]||S[table-format=3.2]S[table-format=1.2]}
               \hline       &\multicolumn{2}{c}{\bf transportable } & \multicolumn{2}{c}{\bf stationary }\\
               {\bf effect} &\multicolumn{1}{c}{\bf corr.} &  \multicolumn{1}{c}{\bf unc.}& \multicolumn{1}{c}{\bf corr.} & \multicolumn{1}{c}{\bf unc.} \\ 
               \hline
               \hline
         BBR ac Stark shift           & 488.3 & 0.9   &492.4 & 2.3\\
	       BBR oven  	                  & 0	    & 0	    &0.9   & 0.9\\
	       Scalar/tensor lattice shift  &-5.7   & 6.1   &-0.3  & 0.9\\
	       higher order lattice shifts  &-0.2   & 0.3   &-0.6  & 0.3\\
	       probe light Stark shift      & 0     & 0.03  &0     & 0.002\\
         cold collisions    	        & 1.6   & 4.1   &0     & 0.1\\
	       backgr. gas collisions	      &-0.5   & 0.5   &-0.2   & 0.2\\
%         line  pulling 		            & 0   	& 0.02	&0 	    & 0.0004\\
         2$^{\rm nd}$ order  Zeeman shift  & 10.9  & 0.5   &3.3   & 0.1\\
         servo error		              & 0     & 0.03   &0 	   & 0.05\\
%	       optical path length	        & 0	    & 0.1 	&0    	& 0.000003\\
	       tunneling		                & 0	    & 0	    &0 	    & 0.1\\
	       DC Stark shift		            & 0     & 0.03  &0	    & 0.03\\
				       \hline
	       {\bf total} 		              & 494.4 & 7.4   &495.6  & 2.6\\
               \hline
	       %height difference	    &-0.9   & 0.06  &0      & 0\\
	\end{tabular}
	\caption{\label{tab:uncertainty}Corrections (corr.) and uncertainties (unc.) of the transportable and stationary clocks in parts of $10^{-17}$.}
\end{table}

During a first set of ratio measurements, the 813~nm light for the optical lattice has been delivered by a diode laser and amplified with a tapered amplifier (TA) chip.
Such laser systems are known to cause problems in the determination of the ac-Stark shift cancellation wavelength due to spectral impurities caused by amplified spontaneous emission \cite{let12}.
Nevertheless, it has been chosen for its compactness and mechanical robustness.
The light from the TA~laser system is spectrally filtered by a volume Bragg grating with a bandwidth of 0.1~nm.
The filtered beam is sent through a 1~m long, large-mode-area (LMA) fiber to remove spatial and spectral correlations that could have been introduced by the grating.
After this fiber, an AOM serves for power stabilization before the light is delivered through a second LMA fiber to the physics package.
However, the comparison of both Sr lattice clocks has revealed a fractional frequency difference of about $3 \times 10^{-16}$ that is not compatible with the combined uncertainty of the clocks of below $1 \times 10^{-16}$.
In addition, the observed Stark shift cancellation frequency has been lower by 101~MHz
%Originfile15-11-26\comb_quick
than expected for the parallel alignment of the lattice polarization vector and the bias field, which even falls outside the wavelength range expected for any lattice polarization \cite{wes11}.

We have replaced the TA diode laser by an titanium sapphire laser and repeated the measurements.
After reevaluation of the TOC we found $\nu_{\rm trans}/\nu_{\rm stat} - 1 = -6(80) \times 10^{-18}$ including a redshift correction of $-9.0(6)\times10^{-18}$.
Here, we use the extrapolated observed instability (Fig.~\ref{fig:lw_ro_Allan}) at the full length of the data set of $1.5 \times 10^{-17}$ as statistical uncertainty.
%derivation in SR1-SR2_comparison_with_lattice@35MHz.xlsx
%essentially statistically weighted average AFTER corrections. No weighting due to systematic uncertainty (very much the same)
%
%NOT TRUSTED whatever is below, not sure whether corrections have been taken into account and corrections have altered
%based on data from 17.12.2015, Sr1 at MOT 
% 04: 11.8 mHz, 1904 points
% 07 10: -4.6 mHz, 6068 points
% wheighted average 0.7 mHz = 1.6E-18
% stat: extrapol to full dataset length: 1.3E-15/sqrt(1904+6068) = 1.45E-17
% syst: Sr1: used files are 04, 07, 10. 10 has small uncertainty (1.6E-17), but is just a quarter of all, thus 07 (approx = 04): 2.6E-17
%				Sr2: 6.9e-17 
The complete uncertainty budgets of both clocks are listed in Tab.~\ref{tab:uncertainty}.

%Meanwhile, this lattice laser system has been replaced by a robust Ti:Sapphire laser of the same type as used in the stationary clock, which allows for transportation with only minor realignment.
%Furthermore, the apparatus has been moved from the laboratory into its container (Fig.~\ref{fig:container}). 

In conclusion, we have built and characterized a TOC that achieves a systematic uncertainty of $7.4 \times 10^{-17}$ and an instability  of $1.3 \times 10^{-15}\prt$.
Note that the gross of the uncertainty stems from the uncertainty in the lattice light shift.
This will be significantly reduced by a full characterization of the new lattice laser system that has been limited due to the short measurement time available.
Therefore, we expect the BBR related uncertainty to become the limiting uncertainty in the near future, which is already below $1 \times 10^{-17}$ and can be reduced further by dedicated probes as presented in \cite{blo14}.
An interrogation laser with improved frequency stability is under development and will reduce the instability of the clock.
Already now, our TOC is outperforming the best transportable Cs fountain clock by one order of magnitude in systematic uncertainty and two orders in instability.
Compared to the recently reported performance of the transportable Ca-ion clock \cite{cao16}, our clock takes about 300-fold less averaging time to reach any given statistical uncertainty.
The TOC  is in a transportable container ready for applications like chronometric leveling, international clock comparisons and precision measurements for fundamental physics.

This work was supported by QUEST, DFG through the RTG~1729 and CRC 1128~geo-Q, the Marie-Curie Action ITN FACT, and the EMRP project ITOC. The EMRP is jointly funded by the EMRP participating countries within EURAMET and the European Union.

%\begin{thebibliography}{10}
	
%\bibliography{C:/Moria/Arbeit/Papers/texbib/texbi431}
%\bibliography{texbi431}
\bibliography{O:/4-3/4-3-Alle/Papers/TeXBib/texbi431}
%\end{thebibliography}
\end{document}